\documentclass[12pt,aps,prd]{revtex4}
\usepackage{epsfig}

\newcommand{\AmS}{{\protect\the\textfont2
  A\kern-.1667em\lower.5ex\hbox{M}\kern-.125emS}}
\def\lQ{\Lambda_{\rm QCD}}

\def\siml{{\
    \lower-1.2pt\vbox{\hbox{\rlap{$<$}\lower6pt\vbox{\hbox{$\sim$}}}}\ }} 

\begin{document}

\title{Heavy Quarkonium and QCD Nonrelativistic  Effective Field Theories
\footnote{Invited Talk given at the ``Workshop on Charmonium Spectroscopy: Past and Future'', Genova (Italy),
June 7-8, 2001.}} 
\author{Nora Brambilla}
\email{nora.brambilla@mi.infn.it}
\affiliation{Dipartimento di Fisica, Via Celoria 16, 20133, Milano, Italy}
\begin{abstract}
QCD nonrelativistic effective field theories (NREFT) are  the modern and most suitable 
frame to describe  heavy quarkonium properties. In this talk  I summarize 
few relevant concepts and some interesting physical applications of NREFT.
\end{abstract}

\maketitle

\section{INTRODUCTION}
\label{sec:intro}
Heavy quarkonium systems play a key role in  a large range of ongoing or planned experiments,
from the search of hybrids to the quarkonium production, from the quark-gluon plasma formation to 
the next linear  collider physics. Being nonrelativistic systems, they enjoy a degree 
of simplification with respect to the other quark bound systems and thus appear to be 
the most appropriate laboratory to study the confinement mechanism and the QCD vacuum 
properties.  Being bound systems with a characteristic radius $r$  extending from the short 
range ($r < 1/\Lambda_{\rm QCD}$) to the long range ($r > 1/\Lambda_{\rm QCD}$) they probe 
the transition region from the perturbative to the nonperturbative regime.
For all these reasons,  it is relevant to be able to treat these systems 
inside QCD and in a model independent approach. In particular, to relate 
the properties of heavy quarkonium directly to the fundamental QCD parameters,
like $\alpha_s$ and $\Lambda_{\rm QCD}$. \par
 The study of heavy quark-antiquark systems is an old topic, see \cite{lectures}
 for some reviews. The spectra show that the gap between the energy levels of such systems is 
much smaller then the mass of the constituent quarks. Thus, they are nonrelativistic systems and 
can be described in first approximation by a Schr\"odinger equation with a potential. 
The form of the potential  may be phenomenologically constrained by the structure of the energy levels
which points to  something  intermediate between  a Coulomb and a harmonic oscillator  potential.
A whole zoo of phenomenologically inspired potentials has been used in the past to reproduce/predict the properties of the spectra. The main ingredients of such potentials remain a Coulomb term superimposed with a linear
(confining) term \cite{lectures}.   But in spite of their success \cite{eichten,rich},
 the limitations of such phenomenological potential models are clear: the theoretical understanding is poor, there is no room for systematic improvement, there is no 
clear relation between the parameters of the potential models and the fundamental parameters of QCD.
In such approaches the confining interaction is imposed by hand.
Thus, relativistic corrections to the static potential are added typically in a complete model dependent way
and there is no systematic procedure to take into account retardation effects which are typically related to 
low energy gluons. This last thing becomes  particularly relevant in QCD where nonperturbative contributions may  
appear also as nonpotential effects (carried e.g. by the gluon condensate \cite{leut}). This 
lead in the past to many inconsistencies and contradictory statements  about the 
existence or not of the $q\bar{q}$ potential. \par
A more rigorous method to obtain the potentials from QCD was developed inside the Wilson loop approach
where the potentials are calculated as vacuum expectation values of Wilson loops and (chromo)electric and 
 (chromo)magnetic field insertions inside Wilson loops \cite{lectures,wilson}. Such objects 
are suitable for a direct lattice evaluation or a calculation inside QCD vacuum models. However, also 
this approach is missing part of the dynamics
 that may characterize a particular heavy quarkonium state,
like some of the short distance higher order contributions and nonpotential effects.\par
It appears that in the several approaches to quarkonium physics there is always something out of control.
This is related, on one hand to the many physical scales that enter the problem and control 
several dynamic effects: perturbative (hard scale) and nonperturbative (low scale) effects; potential and nonpotential
contributions, local and nonlocal condensates contributions. On the other hand,  
it is due to the lack of a fully systematic approach 
endowed with  a unambiguous power counting (in some small parameter) that allows to estimate clearly 
the order of magnitude, and thus the relevance, of the neglected contributions.
The effective field theory (EFT) approach satisfactory eliminates such difficulties \cite{reveff,nrqcd,eff,pnrqcd}. 
Indeed, the existence of a hierarchy of energy scales in quarkonium systems
 allows the construction of EFT with less and less degrees of freedom. This 
leads ultimately  to a field theory derived quantum mechanical description of these systems. 
We call pNRQCD 
 (potential Non Relativistic QCD)
the corresponding EFT \cite{pnrqcd,eff,reveff}. It is important to stress that all the EFT that we will introduce here are, by construction,
completely equivalent to QCD. The procedure through  which such an equivalence is imposed and the integration of the 
degrees of freedom  is done in practice, is called ``matching''
 \cite{reveff}.\par
To be able to disentangle the scales of the bound state, as the EFT
allows us to do, is of key importance in QCD, where we have a confinement region and we would like to be able 
to 'factorize' as much as possible the high energy physics from the low energy physics, dominated by 
nonperturbative effects. Even inside a pure lattice approach,  we have to resort to 
the EFT approach in order to be able  to eliminate the non relevant scales and thus make   
the heavy quarkonium system fits  inside the present capabilities of lattice QCD \cite{latticenrqcd}.\par
Due to space limitations, the present paper is only    a guided  (and partial) collection of 
recent results and references in QCD NREFT for heavy quarkonium. 
The reader is warmly suggested  to refer to the quoted papers for 
all the details and the explanations.

\section{Effective Field Theories for Heavy Quarkonium}
\label{sec:leng}
Being nonrelativistic, heavy quarkonium systems are characterized by, at least, three widely separated scales: 
the mass $m$ of the quark, the (soft) scale associated with the relative momentum $p \sim mv$, $v\ll 1$, and 
the (ultrasoft) scale associated with the typical kinetic energy $E\sim mv^2$. 
Also the inverse of the typical size of the system $r^{-1}$  is of order $mv$.
Here $v$ is the velocity of the quark 
in the bound system and what matters for the following is only that $v$ is a small number. The power counting 
of the EFT will be established in powers of $v$. This point requires special care in the case of 
 charmonium where it is not  clear  if $v$ remains a sufficiently small number.
Moreover, in dependence of the specific system, the scale of the nonperturbative physics $\Lambda_{\rm QCD}$, may turn out to be close to some of the above dynamical scales. The physical picture, which then arises, may be quite different from the perturbative situation. What remains true for all heavy quarkonia is that $m \gg \Lambda_{\rm QCD}$  and thus 
at least the mass scale can be treated perturbatively, i.e. integrated out from QCD order by order 
in the coupling constant $\alpha_s$. The resulting EFT is called NRQCD (Non Relativistic QCD) \cite{nrqcd}.
 The Lagrangian of NRQCD can be organized in powers of $1/m$, thus making explicit the non-relativistic 
nature of the physical systems.   In order for an effective field theory to be useful, a power counting is needed.
The power counting of NRQCD (organized in powers of $v$ and $\alpha_s$) follows from arguments valid in the 
perturbative regime (which should correspond strictly speaking to $\Lambda_{\rm QCD} \siml mv^2$)\cite{reveff}.
Moreover, being still two scales (the momentum and energy scales) dynamical, the matrix elements do not have a unique 
power counting beyond leading order \cite{eff}.
NRQCD allows us to calculate on the lattice systems like bottomonium \cite{latticenrqcd}. The new and very 
successful predictions of NRQCD on inclusive quarkonium decays and on quarkonium production are well known \cite{nrqcd}.

\section{Small radius systems and pNRQCD (for $mv \gg \lQ$)}

In NRQCD  still the dominant role of the potential as 
well as the quantum mechanical nature of the problem are not yet maximally exploited. A higher 
degree of simplification may still be achieved.  In other words, we want to build 
another effective theory for the low energy region  of the non-relativistic bound-state, i.e. we want 
an EFT where only the ultrasoft  degrees of freedom remain dynamical, while the unwanted degrees of freedom  
are integrated out.  To this aim we integrate out the scale of the momentum transfer $\sim mv$ which is supposed 
to be the next relevant scale. Then, two different situations may exist. In the first one, $mv > \Lambda_{\rm QCD}$ 
and thus the matching from NRQCD to pNRQCD may be performed in perturbation theory, expanding in $\alpha_s$.
This is the situation that I will discuss in this paragraph. In the second situation, $mv \siml \Lambda_{\rm QCD}$, 
the matching has to be nonperturbative, i.e. no expansion in $\alpha_s$ is allowed. 
Recalling that $r^{-1} \sim mv$, these two situations correspond  to systems with inverse typical radius smaller or 
bigger than $\Lambda_{\rm QCD}$, or systems respectively dominated by the short range or long range (with respect to the confinement radius) physics.       Although no direct measurements of the typical radius is possible, from all the 
information we have at hand we can say that charmonium belongs to the second case and we will discuss it 
together with the nonperturbative matching to pNRQCD in Sec.4.\par
Now, we briefly describe the case in which $ mv >\Lambda_{\rm QCD}$. 
 At the scale of the matching $\mu^\prime$ 
($mv \gg \mu^\prime \gg mv^2, \lQ$) we have still quarks and gluons.
The effective degrees of freedom are: $Q\bar{Q}$ states (that can be decomposed into 
a singlet  and an octet wave function
under color transformations) with energy of order of the next relevant 
scale, $O(\Lambda_{\rm QCD},mv^2)$ and momentum   ${\bf p}$ of order $O(mv)$,  plus 
ultrasoft gluons $A_\mu({\bf R},t)$ with energy 
and momentum of order  $O(\lQ,mv^2)$. All the  gluon fields are multipole 
expanded (i.e.  in $r$). The Lagrangian is then  an expansion 
in the small quantities  $ {p/m}$, ${ 1/r  m}$ and in   
$O(\Lambda_{\rm QCD}, m v^2)\times r$.

The EFT we obtain \cite{pnrqcd,eff,reveff} produces a zero order equation and 
correction  interactions terms  of the type \cite{pnrqcd,reveff}
\begin{eqnarray*}
\,\left.
\begin{array}{ll}
&
\displaystyle{\left(i\partial_0-{{\bf p}^2 \over 2m}-V_0(r)\right)\Phi({\bf r})=0}
\displaystyle{\quad \quad \quad +  {\rm  corrections\; to\; the\; potential}} \nonumber
\\
&
\displaystyle{+{\rm interaction \;with\; other\; low-energy\; degrees \;of\; freedom}}
\end{array} \right\} 
{\rm pNRQCD}
\end{eqnarray*}
where $V_0(r)=-C_f\alpha_s/r$ in the perturbative tree level case for the singlet and $\Phi({\bf r})$ is the
${\bar Q}$--$Q$ (singlet or octet) wave-function.

The equivalence of pNRQCD to NRQCD, and hence to QCD, is enforced 
by requiring the Green functions of both effective theories to be equal (matching).
In practice, appropriate off shell amplitudes are compared in NRQCD and in pNRQCD,
order by order in the expansion in $1/m$, $\alpha_s$  and in the multipole expansion.
The difference is encoded in  potential-like matching coefficients 
that depend non-analytically on the scale that has been integrated out (in this case ${\bf r})$.\par
At the leading order (LO) in the multipole expansion, 
the equations of motion of the singlet field is the 
 Schr\"odinger equation. 
Therefore pNRQCD has made explicit the dominant role of the potential 
and the quantum mechanical nature 
of the bound state. In particular both the kinetic energy and the potential 
count as $mv^2$ in the $v$ power counting.
The actual bound state  calculation turns out to be very similar to a standard 
quantum mechanical calculation, the only difference being that the wave 
function field couples to US gluons in a field theoretical fashion. 
 From the solution of the Schr\"odinger equation we obtain  the leading order propagators for the singlet and the 
octet state, while the vertexes come from the interaction terms at the NLO (next-to-leading order) 
in the multipole expansion.  In fact the pNRQCD Lagrangian
contains  retardation (or nonpotential) effects that 
start at the NLO in the multipole expansion. 
Thus, pNRQCD has explicit potential terms and thus   it embraces a description of heavy quarkonium 
in terms of potentials. However, it has also  explicit dynamical ultrasoft gluons  and thus  it describes nonpotential 
(retardation) effects. As we see,  such an effective theory is able to provide a solution to the problems
mentioned in the introduction.

The power counting is unambiguous. Calculations can be performed  systematically in the $v$ expansion
and can be improved at the desired order. 
Perturbative (high energy) and nonperturbative (low energy) contributions are disentangled.
Renormalization group improvement  may be performed in the effective theory\cite{rg}.\par
This allows us to  systematically parameterize the nonperturbative contributions that we are not able 
to evaluate directly. 
There are two main situations. If $mv^2 \leq \Lambda_{\rm QCD}$ 
the system is described up to order $\alpha_s^4$ by a potential
 entirely accessible to perturbative QCD. Nonpotential effects start at order $\alpha_s^5 \ln \mu^\prime$
and have been calculated in \cite{nnll}.  We call Coulombic this kind of systems.
Nonperturbative effects are of non potential type and can encoded into 
local (a la Voloshin-Leutwyler) or nonlocal condensates. 
In the second case, when  $mv \gg \Lambda_{\rm QCD} \gg mv^2$, nonperturbative 
contributions to the potential arise when integrating out the scale $\Lambda_{\rm QCD} $ \cite{pnrqcd}. 
We call quasi-Coulombic the systems where the nonperturbative  piece of the potential can be considered small 
with respect to the Coulombic one. 
Some levels of  toponium, the lowest level of $b \bar{b}$ may be considered Coulombic systems, while the 
$J/\psi$, the $\eta_c$, the lowest level of $B_c$ and part of the bottomonium excited levels maybe considered as 
quasi-Coulombic.  Detailed calculation of the properties of such systems in this frame may be found in 
\cite{bott,bc,tt}. In particular in \cite{bott} an accurate  determination of the mass of the $b$ is also obtained.
\par
As it is typical in an effective theory, only the actual calculation may confirm if the initial assumption  about the physical system was appropriate. 

\section{Charmonium and the nonperturbative matching to pNRQCD}

With the exception of the lowest state, which maybe a quasi-Coulombic system, the main part of the 
excited levels of charmonium probe the confinement region $mv\siml \Lambda_{\rm QCD}$. 
Then, pNRQCD should be obtained 
via a nonperturbative matching, i.e. without expansions in $\alpha_s$.
This have been proved to be equivalent to compute the heavy quarkonium potential at order $1/m^2$ \cite{nonp}.
More precisely, a pure potential picture emerges at the leading order in the ultrasoft expansion under the condition that all the gluonic excitations (hybrids) have a gap of order $\Lambda_{\rm QCD}$ \cite{nonp}.  
Higher order effects in the $1/m$ expansion as well as extra ultrasoft degrees of freedom such as hybrids or pions can be systematically included and may eventually affect  the leading potential picture.
Thus we recover the quark model from pNRQCD\cite{nonp}. 
The final result for the potentials (static and relativistic corrections) appear factorized in a part 
containing the high energy  dynamics (and calculable in perturbation theory) which is inherited from NRQCD,
and a part containing the low energy dynamics given in terms of Wilson loops and chromo-electric and chromo-magnetic 
insertions in the Wilson loop \cite{nonp,reveff,lectures}. 
Such low energy contributions can be simply calculated  on the lattice 
or evaluated in QCD vacuum models. Also in this case the power counting supply us with a valuable and systematic  way 
of estimating the size of the neglected terms.
Moreover, since the power counting of pNRQCD may be different from the power counting of NRQCD, we expect 
that we may eventually explain  in this way the difficulties that NRQCD is facing in explaining the polarization of the prompt $J/\psi$ data \cite{prod}.
New and quite interesting predictions have been obtained in this frame also for charmonium 
 $P$ wave inclusive decays \cite{decay} (see also\cite{mussa})
and  on the behaviour of the heavy (and thus also charmed) hybrids  potential for small $r$ \cite{pnrqcd}.

\section{Conclusions}
An effective theory of QCD which describes heavy quarkonium has been constructed systematically 
and within a controlled expansion. Such a theory disentangles the scales of the bound state and has 
a definite power counting. All known perturbative and nonperturbative, potential and nonpotential 
regimes are present and separately factorized in  the theory.  In this way the properties of heavy quarkonium 
and in particular  of charmonium, 
which   is the subject of the present  conference, are related to the fundamental parameters of QCD.  Such an effective theory is thus the appropriate frame to  perform calculation of heavy quarkonium 
properties. In this  paper  we presented   just  a guided recollection of recent references where 
new results on   heavy quarkonium spectra, decays, production and heavy quarkonium hybrids potentials 
can be found  inside  the frame of pNRQCD.
\vspace{0.8cm}

\noindent{\bf Acknowledgments} I would like to thank Claudia Patrignani  and the other organizers for the invitation 
and for the perfect organization of this very interesting workshop. I thank the Alexander Von Humboldt 
foundation for support.

\end{document}